\PassOptionsToPackage{unicode}{hyperref}
\PassOptionsToPackage{hyphens}{url}
\documentclass[
]{article}
\usepackage{amsmath,amssymb}
\usepackage{lmodern}
\usepackage{iftex}
\ifPDFTeX
  \usepackage[T1]{fontenc}
  \usepackage[utf8]{inputenc}
  \usepackage{textcomp} 
\else 
  \usepackage{unicode-math}
  \defaultfontfeatures{Scale=MatchLowercase}
  \defaultfontfeatures[\rmfamily]{Ligatures=TeX,Scale=1}
\fi
\IfFileExists{upquote.sty}{\usepackage{upquote}}{}
\IfFileExists{microtype.sty}{
  \usepackage[]{microtype}
  \UseMicrotypeSet[protrusion]{basicmath} 
}{}
\makeatletter
\@ifundefined{KOMAClassName}{
  \IfFileExists{parskip.sty}{%
    \usepackage{parskip}
  }{
    \setlength{\parindent}{0pt}
    \setlength{\parskip}{6pt plus 2pt minus 1pt}}
}{
  \KOMAoptions{parskip=half}}
\makeatother
\usepackage{xcolor}
\IfFileExists{xurl.sty}{\usepackage{xurl}}{} 
\IfFileExists{bookmark.sty}{\usepackage{bookmark}}{\usepackage{hyperref}}
\hypersetup{
  hidelinks,
  pdfcreator={LaTeX via pandoc}}
\usepackage{longtable,booktabs,array}
\usepackage{tabularx}
\usepackage{multirow}
\usepackage{makecell}
\usepackage{ragged2e}
\usepackage{calc} 
\usepackage{etoolbox}
\makeatletter
\patchcmd\longtable{\par}{\if@noskipsec\mbox{}\fi\par}{}{}
\makeatother
\IfFileExists{footnotehyper.sty}{\usepackage{footnotehyper}}{\usepackage{footnote}}
\makesavenoteenv{longtable}
\usepackage{graphicx}
\makeatletter
\def\maxwidth{\ifdim\Gin@nat@width>\linewidth\linewidth\else\Gin@nat@width\fi}
\def\maxheight{\ifdim\Gin@nat@height>\textheight\textheight\else\Gin@nat@height\fi}
\makeatother
\setkeys{Gin}{width=\maxwidth,height=\maxheight,keepaspectratio}
\makeatletter
\def\fps@figure{htbp}
\makeatother
\ifLuaTeX
  \usepackage{selnolig}  
\fi

\author{}
\date{}

\begin{document}

A Kubernetes custom scheduler based on reinforcement learning for compute‑intensive pods

HANLIN ZHOU

\href{https://cs.usm.my/}{School of Computer Sciences,~Universiti Sains Malaysia}

\href{https://cs.usm.my/}{Xiamen Institute of Software Technology, China;}

zhouhanlin1@student.usm.my

HUAH YONG CHAN

\href{https://cs.usm.my/}{School of Computer Sciences,~Universiti Sains Malaysia}

\href{mailto:hychan@usm.my}{\nolinkurl{hychan@usm.my}}

SHUN YAO ZHANG

Xiamen Institute of Software Technology, China;

zhangshunyaogong@sina.com

MEI E LIN

\href{https://cs.usm.my/}{Xiamen Institute of Software Technology, China;}

\href{mailto:hychan@usm.my}{linmeie@xmist}.edu.cn

JINGFEI NI

Information Management Department of Manzhouli Customs, China

morgan\_ni@163.com

With the rise of cloud computing and lightweight containers, Docker has emerged as a leading technology for rapid service deployment, with Kubernetes responsible for pod orchestration. However, for compute‑intensive workloads---particularly web services executing containerized machine‑learning training---the default Kubernetes scheduler does not always achieve optimal placement. To address this, we propose two custom, reinforcement‑learning--based schedulers, SDQN and SDQN‑n, both built on the Deep Q‑Network (DQN) framework. In compute‑intensive scenarios, these models outperform the default Kubernetes scheduler as well as Transformer‑ and LSTM‑based alternatives, reducing average CPU utilization per cluster node by 10\%, and by over 20\% when using SDQN‑n. Moreover, our results show that SDQN‑n's approach of consolidating pods onto fewer nodes further amplifies resource savings and helps advance greener, more energy‑efficient data centers.Therefore, pod scheduling must employ different strategies tailored to each scenario in order to achieve better performance.Since the reinforcement‑learning components of the SDQN and SDQN‑n architectures proposed in this paper can be easily tuned by adjusting their parameters, they can accommodate the requirements of various future scenarios.

\textbf{CCS CONCEPTS} • Cloud Computing Technologies • Machine Learning

\textbf{Additional Keywords and Phrases:} Kubernetes, Scheduler, Reinforcement‑learning, Machine Learning

\hypertarget{introduction}{%
\section{\texorpdfstring{\textbf{1}Introduction}{1Introduction}}\label{introduction}}

Over the past decade, cloud computing and virtualization technologies have advanced at an unprecedented pace, driven in large part by the emergence of lightweight containerization solutions such as Docker. Owing to its rapid deployment capabilities, Docker has been widely adopted as the preferred infrastructure for business-critical services{[}\href{/l}{1}{]}. Containerized environments are commonly employed to launch NGINX instances for high‑concurrency web services or to execute compute‑intensive workloads like scientific simulations. In the realm of deep learning, Docker ensures consistent software environments across experiments, thereby enhancing result reproducibility. In real‑world deployments, however, fluctuating demand leads to continual variation in the number of pods required to sustain service levels.

In order to achieve effective container orchestration and management, tools such as Kubernetes are required. As one of the most widely adopted orchestration platforms in recent years, Kubernetes provides not only container scheduling and management but also features such as load balancing, automatic scaling, and self‑healing. For many organizations, the ability to scale containers ensures smooth transitions through periods of high and low demand, thereby maintaining high availability. However, beyond the Horizontal Pod Autoscaler (HPA){[}\href{/l}{2}{]}, container counts continue to fluctuate in response to business requirements. Container increases arise for two reasons: first, activation of HPA, which automatically adjusts pod counts based on current CPU or memory utilization---when thresholds are reached, additional pods are provisioned; and second, business‑driven creation of large numbers of pods, either manually or via automated processes.

By default, Kubernetes employs a built‑in scheduling mechanism. Scheduling in Kubernetes involves placing pods onto appropriate nodes so that the node's Kubelet can execute them. The default scheduler,\,kube‑scheduler, operates in two stages---filtering and scoring---evaluating candidate nodes through a series of functions that consider factors such as hardware, software, and affinity/anti‑affinity rules. While the default scheduler serves well in most scenarios, it may yield suboptimal placements under certain workloads, prompting investigation into which scheduling algorithms can maximize resource efficiency.

In both high‑concurrency HTTP web services and scientific computing scenarios, workloads frequently exhibit intensive computational demands, rendering the corresponding pods compute‑intensive. Such workloads also include distributed computing containers and small‑scale, CPU‑based machine‑learning training tasks. Compute‑intensive pods typically execute in short, bursty batches, since they perform rapid, high‑volume calculations before resource consumption drops off or the pod terminates.

This study focuses on compute‑intensive pods and evaluates four Kubernetes scheduling strategies: two reinforcement‑learning--based schedulers (SDQN and SDQN‑n, both derived from the Deep Q‑Network framework) and two neural approaches based on LSTM and Transformer architectures. The objective is to determine which algorithm minimizes average CPU utilization during bulk pod deployment. Average per‑node CPU utilization was selected as the key performance metric because it directly informs CPU provisioning decisions for both cloud (e.g., Alibaba Cloud, AWS) and on‑premises servers, and because high CPU utilization can degrade co‑located services, increase power consumption, and limit the ability to decommission under‑utilized nodes.

Experimental results demonstrate that, in targeted compute‑intensive scenarios, the RL‑based schedulers outperform the default Kubernetes scheduler: SDQN reduces average CPU utilization by 10\%, while SDQN‑n achieves reductions exceeding 20\%. Furthermore, both SDQN and SDQN‑n yield lower CPU utilization than the LSTM‑ and Transformer‑based schedulers, indicating that reinforcement‑learning--driven scheduling offers superior cluster resource savings compared to both default and alternative AI‑driven approaches.

Contributions of this paper

\begin{enumerate}
\def\labelenumi{\arabic{enumi}.}
\item
  Reinforcement‑Learning--Enhanced Scheduling: Introduction of the SDQN framework, which combines the Deep Q‑Network (DQN) reinforcement‑learning paradigm with Kubernetes's scheduling pipeline, and demonstration that, in batch deployments of compute‑intensive pods, SDQN achieves lower cluster‑wide CPU utilization than both the default scheduler and Transformer‑ or LSTM‑based alternatives.
\end{enumerate}

2.RL‑Driven Pod Consolidation: Development of the SDQN‑n strategy, which integrates reinforcement‑learning--based decision‑making with a pod‑consolidation approach---concentrating compute‑intensive pods onto a subset of nodes---to further reduce overall CPU usage by over 20\%, thereby enabling the shutdown of idle machines and advancing greener, more energy‑efficient data centers{[}\href{/l}{3}{]}.

The remainder of the paper is organized as follows: Section\,2 reviews related work, Section\,3 presents the theoretical foundations, Section\,4 details the implementation, Section\,5 analyzes the experimental data, and Section\,6 concludes the study.

\hypertarget{related-work}{%
\subsection{\texorpdfstring{2 Related Work }{2 Related Work }}\label{related-work}}

There are three types of resource scaling in Kubernetes: HPA, VPA (Vertical Pod Autoscaler), and CA (Cluster Autoscaler). By default, HPA adjusts the number of pods based on resource usage such as CPU and memory{[}\href{/l}{3}{]}. However, sometimes CPU or memory usage may not accurately reflect the service status of the corresponding pod. For instance, in Nginx containers aimed at responding to user HTTP requests{[}\href{/l}{4}{]}, if there is a service anomaly due to reasons like program crashes or upstream service issues, the CPU and memory usage of the container itself or its upstream services may still appear normal{[}\href{/l}{5}{]}, but users may receive a large number of 5xx status codes. This leads to two issues when using HPA with default metrics. First, HPA cannot scale itself based on the values of the pod. Second, HPA only has CPU and memory as default metrics, and using custom metrics requires the installation of an additional API server.Current research on Kubernetes scheduling can be grouped into three primary areas. First, parameter‑specific scheduler optimizations---such as network‑bandwidth‑aware and communication‑latency‑aware schedulers; Second, IoT and edge‑cloud‑centric scheduling enhancements; and Third, machine‑learning--driven scheduler designs.

Chiaro et al.\,{[}\href{/l}{6}{]} propose a multi‑cluster Kubernetes scheduler optimized for end‑to‑end latency to improve Quality of Experience, while Dehnashi et al. {[}\href{/l}{7}{]} co‑locate high‑communication pods within the same cluster and apply a node‑selection algorithm that yields a 2.62\(\times\) performance improvement over the default scheduler. Marchese et al.\,{[}\href{/l}{8}{]} introduce an SLO‑aware scheduling and descheduling strategy that maintains response‑time targets under unstable network conditions.

Beyond network‑stability optimizations, Wang et al.\,{[}\href{/l}{9}{]} present an Edge Information--Aware Scheduler (EIS) that leverages edge‑cluster topology and performance metrics to place containerized IoT applications, reducing network latency by 18\%. Saito et al.\,{[}\href{/l}{10}{]} model the hierarchical structure of the edge‑cloud continuum to inform Kubernetes microservice scheduling, achieving significant end‑to‑end response‑time reductions. Park et al.\,{[}\href{/l}{11}{]} design an accelerator‑information extraction module to enhance edge‑computing scheduler decisions, and Qiao et al.\,{[}\href{/l}{12}{]} develop a TOPSIS‑based Network‑Aware Container Scheduling (NACS) algorithm that increases web application throughput by 26\% and Redis throughput by 29.6\%.

In the realm of machine‑learning--driven scheduling, Jian et al.\,{[}\href{/l}{13}{]} introduce DRS, a deep‑reinforcement‑learning scheduler that monitors multiple parameters and delivers a 27.29\% improvement in resource utilization with only 3.27\% CPU overhead and 0.648\% communication‑latency overhead compared to kube‑scheduler, while reducing load imbalance by 2.90\(\times\). Jorge‑Martinez et al.\,{[}\href{/l}{14}{]} propose the AI‑based Kubernetes Container Scheduling Strategy (KCSS) to guide load migration and improve scheduling efficiency. Wang et al.\,{[}\href{/l}{15}{]} present PPO‑LRT, a Proximal Policy Optimization--based RL approach that adaptively dispatches edge tasks to minimize response time, achieving an average reduction of approximately 31\%. Cheng et al.\,{[}\href{/l}{16}{]} develop CSFRL, a fine‑grained compute‑capability scheduling framework using ranking‑based PPO, and Rothman et al.\,{[}\href{/l}{17}{]} introduce RLKube, a custom scheduler plugin employing DDQN with Prioritized Experience Replay to optimize throughput and energy consumption. The consolidated results are summarized in Table\,1.

Despite these advances, most studies address general‑purpose or network/edge‑focused scenarios; none specifically target scheduling algorithms for compute‑intensive pods, therefore, evaluates the optimization effects of reinforcement‑learning--based schedulers on compute‑intensive pod workloads.

\textbf{Table 1:} Summary of Current Research Directions in Kubernetes Scheduling

\begin{table}[htbp]
\centering
\footnotesize
\begin{tabularx}{\linewidth}{@{}>{\raggedright\arraybackslash}p{0.19\linewidth} >{\raggedright\arraybackslash}p{0.18\linewidth} X X@{}}
\toprule
\textbf{Proposer(s)} & \textbf{Research Direction} & \textbf{Techniques \& Algorithms} & \textbf{Key Outcomes} \\
\midrule
Chiaro [6], Dehnashi[7], Marchese [8] & Parameter-Specific Scheduling & Multi-cluster latency optimization; high-communication pod co-placement; SLO-aware scheduling/descheduling & 2.62$\times$ perf. gain; maintained response-time SLO under unstable networks \\
Wang [9], Saito [10], Park [11], Qiao [12] & IoT/Edge-Cloud-Aware Scheduling & Edge Information--Aware Scheduler (EIS); hierarchical continuum modeling; accelerator info extraction; TOPSIS-based NACSg & -18\% network latency $\downarrow$; Web throughput +26\%; Redis throughput +29.6\% \\
Jian[13], Jorge-Martinez[14], Wang[15], Cheng[16], Rothman  [17] & ML-Driven Reinforcement-Learning & Deep-RL scheduler (DRS); AI-based load control (KCSS); PPO-LRT; ranking-based PPO (CSFRL); DDQN + PER plugin (RLKube)Container Scheduling (NACS) & Resource utilization $\uparrow$27.3\%; response time $\downarrow$\textasciitilde{}31\%; optimized throughput and energy usage \\
\bottomrule
\end{tabularx}
\end{table}

\hypertarget{section}{%
\subsection{}\label{section}}

\hypertarget{related-kubernetes-theory}{%
\subsection{3 Related Kubernetes Theory}\label{related-kubernetes-theory}}

This chapter will introduce some processes about kubernetes and its scheduler.

\hypertarget{kubernetes-fundamentals}{%
\subsection{\texorpdfstring{\textbf{3.1} Kubernetes Fundamentals}{3.1 Kubernetes Fundamentals}}\label{kubernetes-fundamentals}}

Kubernetes is a container orchestration platform for Docker and similar technologies, capable of managing physical or virtual servers grouped into a cluster. Each server---referred to as a node---hosts a variable number of pods, which are the smallest deployable units in Kubernetes. A pod typically contains a single container but can encapsulate multiple containers when necessary. The Kubernetes API Server functions as the control planes frontend{[}\href{/l}{4}{]}, issuing RESTful calls to schedule and manage cluster components, and can be extended via custom resources to handle bespoke workloads. Etcd, a distributed key--value store, underlies Kubernetes state persistence, storing cluster data---such as pod statuses, service definitions, and configuration objects---with high availability, strong consistency, and distributed coordination.

\hypertarget{kubescheduler}{%
\subsection{\texorpdfstring{\textbf{3.2} kube‑scheduler}{3.2 kube‑scheduler}}\label{kubescheduler}}

The kube‑scheduler is a control‑plane component responsible for assigning newly created pods to appropriate nodes within a Kubernetes cluster. In a typical deployment, multiple nodes register with the control plane, and the scheduler evaluates each new pod against available nodes to determine the optimal placement. Scheduling decisions account for the pod resource requests, the current resource availability on each node, and policy constraints such as node‑ and pod‑affinity/anti‑affinity, taints and tolerations, and load distribution.

The scheduling algorithm proceeds in two phases as Figure 1:

Filtering (Predicates):Candidate nodes are screened through a set of predicate functions that verify whether each node can satisfy the pod resource and policy requirements. Nodes that fail any predicate are excluded. If no nodes remain, the pod remains unscheduled until resources become available or constraints change.

Scoring (Priorities):\\
Each node that passes the filtering phase is scored by one or more priority functions, which rank nodes according to metrics such as resource balance, affinity preferences, or custom heuristics. The pod is then bound to the node with the highest aggregate score; in the event of a tie, one of the top‑scoring nodes is selected at random{[}\href{/l}{18}{]}.

Beyond the built‑in predicates and priorities, both can be extended or replaced via custom scheduling profiles. Configuration of additional scheduling behaviors---such as custom predicates, priorities, or scheduler plugins---is accomplished by editing the kube‑scheduler configuration file and specifying the desired extension points.

\includegraphics[width=5.04583in,height=1.14792in]{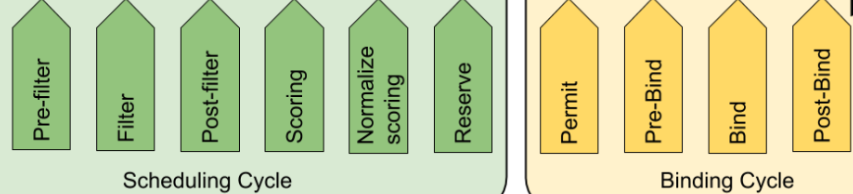}

Figure 1: Kubernetes Default Scheduler Process Diagram

\hypertarget{methodology}{%
\section{\texorpdfstring{\textbf{4} METHODOLOGY}{4 METHODOLOGY}}\label{methodology}}

This study will use the reinforcement‑learning--based schedulers SDQN and SDQN‑n to replace the original default scheduler, so that pods are scheduled using SDQN and SDQN‑n. The overall process is divided into three steps:

\begin{itemize}
\item
  Algorithm Selection and Scheduler Development: Select the appropriate algorithm parameters and implement the custom schedulers SDQN and SDQN‑n.
\item
  Scheduler Replacement: Replace the default scheduler with each of the two custom schedulers in turn.
\item
  Performance Testing: Evaluate the performance of the custom schedulers.
\end{itemize}

\hypertarget{algorithm-selection-of-sdqn-and-sdqn-n}{%
\subsection{\texorpdfstring{\textbf{4.1} Algorithm Selection of SDQN and SDQN-n}{4.1 Algorithm Selection of SDQN and SDQN-n}}\label{algorithm-selection-of-sdqn-and-sdqn-n}}

In the first step, it is necessary to choose the reference parameters, since SDQN uses a neural network to approximate the Q‑function, where Q includes state and action. SDQN is an algorithm based on deep learning and reinforcement learning; Q‑learning is a classical reinforcement‑learning algorithm grounded in the value‑iteration concept, which guides an agent to select the optimal action in each state by estimating the Q‑value for each state‑action pair.

\hypertarget{input-parameters}{%
\subsection{\texorpdfstring{\textbf{4.1.1} Input Parameters}{4.1.1 Input Parameters}}\label{input-parameters}}

Because both machine learning and reinforcement learning require input parameters, the six selected inputs are listed in Table\,2.

Table 2: Input Parameters for Machine Learning and Reinforcement Learning

\begin{table}[htbp]
\centering
\small
\begin{tabularx}{\linewidth}{@{}>{\raggedright\arraybackslash}p{0.28\linewidth} X@{}}
\toprule
\textbf{Parameter Name} & \textbf{Calculation} \\
\midrule
CPU Usage Percentage & (Real-time CPU Usage/CPU Capacity)$\times$100 \\
Memory Usage Percentage & (Real-time Memory Usage/Memory Capacity)$\times$100 \\
Pod Utilization & (Number of Running Pods/Maximum Pods per Node)$\times$100 \\
Health Status & 1 if node status is “Ready,” otherwise 0 \\
Node Uptime (hours) & Difference between current time and node start time (in hours) \\
Number of Running Pods & Absolute count \\
\bottomrule
\end{tabularx}
\end{table}

The six parameters were chosen as experimental inputs because CPU and memory usage constitute essential resource constraints; pod utilization reflects the aggregate workload pressure on each node; health status guarantees that a node is online and functioning correctly; node uptime accounts for the fact that long‑stably running nodes exhibit a lower failure probability than newly added or recently recovered nodes; and the current number of running pods can be used to enforce distribution of workload across a limited set of nodes.

\hypertarget{sdqn-algorithm-and-model}{%
\subsection{\texorpdfstring{\textbf{4.1.2 SDQN Algorithm and Model}}{4.1.2 SDQN Algorithm and Model}}\label{sdqn-algorithm-and-model}}

Table\,3 presents the reward‑function rules employed by the SDQN reinforcement‑learning model.

The reward‑and‑penalty design aims first to maintain CPU and memory utilization within optimal operating ranges, and second to provide a robust scoring mechanism necessary for replacing the default Kubernetes scheduler with SDQN and SDQN‑n.

Table 3: Input Parameters for Machine Learning and Reinforcement Learning of SDQN

\begin{table}[htbp]
\centering
\small
\begin{tabularx}{\linewidth}{@{}>{\raggedright\arraybackslash}p{0.28\linewidth} X@{}}
\toprule
\textbf{Parameter Name} & \textbf{Calculation} \\
\midrule
Initial Base Value & 100 points \\
Health Status & --100 points if node is unhealthy (healthy\_status == 0) \\
CPU Usage & >70\%: --2 points for each 1\% above threshold; 40--70\%: +10 points; otherwise: --10 points \\
Memory Usage & >70\%: --2 points for each 1\% above threshold; 40--70\%: +10 points; otherwise: --10 points \\
Pod Utilization & +20 points if running-pods / max-pods per node $\in$ [0.6, 0.9]; otherwise: --10 points \\
Node Uptime (hours) & $\geq$24 hours: +5 points; otherwise: --5 points \\
Pod Distribution & +5 points for each additional node in the pod distribution \\
\bottomrule
\end{tabularx}
\end{table}

Table\,4 presents the consolidated SDQN model definition and training process.

Table 4: SDQN Model Definition and Training Process

\begin{table}[htbp]
\centering
\small
\begin{tabularx}{\linewidth}{@{}>{\raggedright\arraybackslash}p{0.28\linewidth} X@{}}
\toprule
\textbf{Model Component} & \textbf{Description} \\
\midrule
Input Layer & 6-dimensional input representing the six state features \\
Hidden Layer & Single fully connected layer mapping 6 $\rightarrow$ 32 dimensions; ReLU activation \\
Output Layer & Fully connected layer mapping 32 $\rightarrow$ 1; outputs the estimated Q-value for the input state \\
Loss Function & Mean Squared Error (MSELoss), quantifying the difference between predicted Q-values and target rewards \\
Optimizer & Adam optimizer with a learning rate of 0.001 \\
Training Loop & Forward pass to compute Q(s,a), then backpropagation using target rewards to update network weights \\
\bottomrule
\end{tabularx}
\end{table}

By adopting the above approach, the original default scheduler can be fully replaced, with each reward criterion designed to uphold the normal operational requirements of the nodes.

\hypertarget{sdqn-n-algorithm-and-model}{%
\subsection{\texorpdfstring{\textbf{4.1.3} SDQN-n Algorithm and Model}{4.1.3 SDQN-n Algorithm and Model}}\label{sdqn-n-algorithm-and-model}}

Since SDQN‑n is derived from SDQN by enforcing that newly scheduled pods be placed across a specified number of nodes, while all other conditions remain identical to those of SDQN, Table\,5 highlights only the difference from Table\,3 in the Pod Distribution variable.

Table 5: Input Parameters for Machine Learning and Reinforcement Learning of SDQN-n

\begin{table}[htbp]
\centering
\small
\begin{tabularx}{\linewidth}{@{}>{\raggedright\arraybackslash}p{0.28\linewidth} X@{}}
\toprule
\textbf{Parameter Name} & \textbf{Calculation} \\
\midrule
Pod Distribution(n=2) & \makecell[l]{If the number of candidate nodes $\geq$ 2: placement outside the Top 2: --50 points; placement within the Top 2: +20 points.\\If the number of candidate nodes < 2: if running pods > 0: +20 points; otherwise: --10 points.} \\
\bottomrule
\end{tabularx}
\end{table}

SDQN‑n differs by enforcing the placement of new pods on a specified number of nodes (n\,=\,2). As reflected in the reward criteria, SDQN‑n concentrates new pods onto two nodes whenever possible. Although placement is constrained, node health metrics---such as CPU and memory remaining within healthy ranges---continue to be evaluated; if a chosen node falls outside these thresholds, pods are redirected to other nodes to maintain overall cluster stability.

The SDQN‑n model definition and training procedure mirror those of SDQN in all respects, with the sole distinction being the modified Pod Distribution variable in the reinforcement‑learning component.

\hypertarget{algorithm-selection-of-lstm-and-transformer}{%
\subsection{\texorpdfstring{\textbf{4.2} Algorithm Selection of LSTM and Transformer}{4.2 Algorithm Selection of LSTM and Transformer}}\label{algorithm-selection-of-lstm-and-transformer}}

For comparative evaluation, LSTM and Transformer models were also employed alongside the reinforcement‑learning--based SDQN and SDQN‑n schedulers. The LSTM network architecture is as follows Table 6.

Table 6: LSTM‑Based Custom Scheduler Model

\begin{table}[htbp]
\centering
\small
\begin{tabularx}{\linewidth}{@{}>{\raggedright\arraybackslash}p{0.28\linewidth} X@{}}
\toprule
\textbf{Model Component} & \textbf{Description} \\
\midrule
Input Layer & Six state features, shaped as (1, 1, 6) for a single time step \\
LSTM Layer & Single-layer LSTM with 32 hidden units, employed to extract temporal features \\
Output Layer & Fully connected layer mapping the LSTM’s final time-step output to a single score \\
Loss Function & Mean Squared Error (MSELoss), measuring the discrepancy between predicted scores and target rewards \\
Optimizer & Adam optimizer with a learning rate of 0.001 \\
Training Loop & Forward pass to compute score $\rightarrow$ compute MSE loss $\rightarrow$ backpropagate to update network parameters \\
\bottomrule
\end{tabularx}
\end{table}

\begin{enumerate}
\def\labelenumi{\roman{enumi}.}
\item
  Since the Transformer model is also a widely used machine‑learning architecture, a Transformer‑based custom scheduler with standard parameter settings is likewise employed; the corresponding parameters are presented in Table\,7.
\end{enumerate}

Table 7: Transformer‑Based Custom Scheduler Model

\begin{table}[htbp]
\centering
\small
\begin{tabularx}{\linewidth}{@{}>{\raggedright\arraybackslash}p{0.28\linewidth} X@{}}
\toprule
\textbf{Model Component} & \textbf{Description} \\
\midrule
Input Layer & Six state features, shaped as (1, 1, 6) for a single time step \\
Linear Projection & Projects 6-dimensional input into a 32-dimensional embedding (d\_model = 32) \\
Transformer Encoder & Single-layer encoder with 4 attention heads (nhead = 4) and 1 layer (num\_layers = 1) to extract temporal features \\
Output Layer & Selects the final time-step output from the encoder and applies a fully connected layer to produce a single score \\
Loss Function & Mean Squared Error (MSELoss), measuring the difference between predicted and target scores \\
Optimizer & Adam optimizer with a learning rate of 0.001 \\
Training Procedure & Forward pass to compute score $\rightarrow$ compute MSE loss $\rightarrow$ backpropagation to update model parameters \\
\bottomrule
\end{tabularx}
\end{table}

\hypertarget{simulation-methodology-and-evaluation-metrics}{%
\subsection{\texorpdfstring{\textbf{4.3 Simulation Methodology and Evaluation Metrics}}{4.3 Simulation Methodology and Evaluation Metrics}}\label{simulation-methodology-and-evaluation-metrics}}

\hypertarget{computeintensive-workload-simulation}{%
\subsection{\texorpdfstring{\textbf{4.3.1 Compute‑Intensive Workload Simulation}}{4.3.1 Compute‑Intensive Workload Simulation}}\label{computeintensive-workload-simulation}}

To evaluate reinforcement‑learning--based scheduling for compute‑intensive pods, newly instantiated pods are configured to execute a no‑op, CPU‑bound workload that consumes defined CPU and memory resources without external dependencies, thereby emulating real‑world compute pressure.

\hypertarget{evaluation-metric}{%
\subsection{\texorpdfstring{\textbf{4.3.2 Evaluation Metric}}{4.3.2 Evaluation Metric}}\label{evaluation-metric}}

Scheduling effectiveness is measured by the cluster‑wide average CPU utilization per node, defined as the sum of each node's CPU usage percentage divided by the total number of nodes (including idle ones). This metric captures overall compute‑resource consumption and remaining capacity more directly than latency or memory metrics. For example:

\begin{itemize}
\item
  Uniform distribution: Three nodes each at 20\% CPU \(\rightarrow\) average\,=\,(20\,+\,20\,+\,20)/3\,=\,20\%
\item
  Consolidated distribution: CPU usages of 10\%, 25\%, and 20\% \(\rightarrow\) average\,=\,(10\,+\,25\,+\,20)/3\,\(\approx\)\,18.3\%
\end{itemize}

The lower average in the consolidated case indicates that one node is underutilized (10\%) while another is more heavily loaded (25\%), and one may be idle enough to shut down---demonstrating potential resource savings. Image‑caching and shared I/O further reduce startup overhead for subsequent pods, accentuating these differences. Therefore, average per‑node CPU utilization provides an intuitive, direct measure of scheduler performance under compute‑intensive conditions. Cluster have 4 slave nodes.

\hypertarget{experimental-results-and-analysis}{%
\subsection{\texorpdfstring{\textbf{5 Experimental Results and Analysis}}{5 Experimental Results and Analysis}}\label{experimental-results-and-analysis}}

\hypertarget{experimental-results}{%
\subsection{\texorpdfstring{\textbf{5.1 Experimental Results}}{5.1 Experimental Results}}\label{experimental-results}}

\hypertarget{default-scheduler-experimental-data}{%
\subsection{\texorpdfstring{\textbf{5.1.1 Default Scheduler Experimental Data}}{5.1.1 Default Scheduler Experimental Data}}\label{default-scheduler-experimental-data}}

In this scenario, 50 new pods were instantiated, each executing a no‑op workload to simulate compute‑intensive processing. Figure\,2 presents the results of the first test conducted with the default Kubernetes scheduler.

Test 1, slave1=20, slave2=19, slave3=9, slave4=2. Cluster CPU average utilization 29.27\%

\includegraphics[width=5.30486in,height=2.25903in]{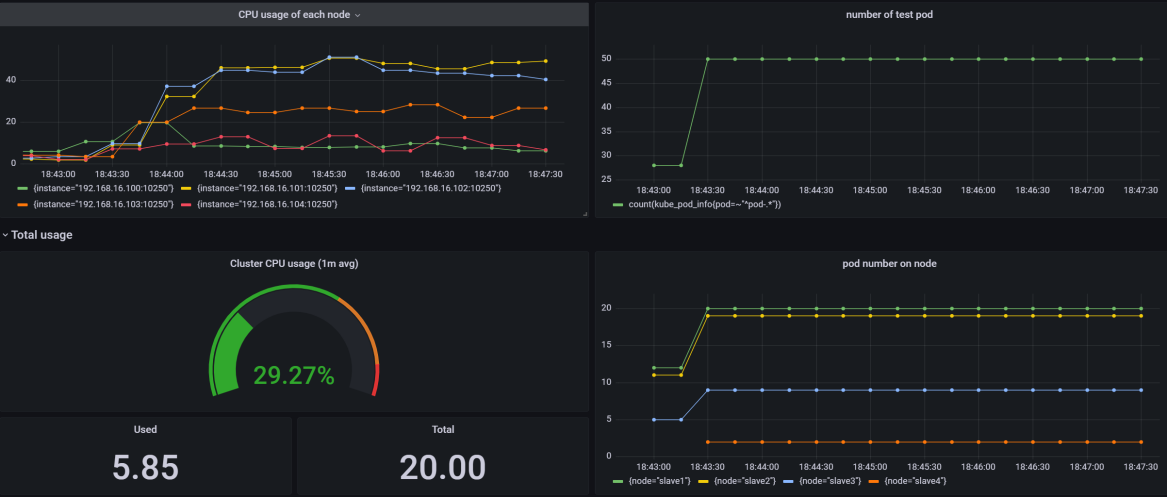}

Figure 2: Default scheduler grafana graph

The above figure is a Grafana example. To guard against the lack of representativeness in a single run, the following presents results from multiple experimental trials as Table 8.

Table 8: Default scheduler multiple data

\begin{table}[htbp]
\centering
\scriptsize
\setlength{\tabcolsep}{3pt}
\begin{tabular}{>{\raggedright\arraybackslash}p{0.20\linewidth} c c c c c c}
\toprule
\textbf{Scheduler Model} & \textbf{Trial} & \multicolumn{4}{c}{\textbf{Pod Distribution}} & \textbf{Average CPU Utilization} \\
\cmidrule(lr){3-6}
 &  & \textbf{Slave 1} & \textbf{Slave 2} & \textbf{Slave 3} & \textbf{Slave 4} &  \\
\midrule
\multirow{5}{*}{Default Scheduler} & 1 & 20 & 19 & 9 & 2 & 29.97\% \\
 & 2 & 20 & 19 & 9 & 2 & 31.82\% \\
 & 3 & 21 & 18 & 9 & 2 & 30.95\% \\
 & 4 & 19 & 12 & 18 & 1 & 29.71\% \\
 & 5 & 19 & 11 & 19 & 1 & 31.91\% \\
\midrule
\textbf{Average} & \multicolumn{5}{c}{Coefficient of Variation (CV) = 2.95\%} & 30.87\% \\
\bottomrule
\end{tabular}
\end{table}

\hypertarget{sdqn-and-sdqn-n-scheduler-experimental-data}{%
\subsection{\texorpdfstring{\textbf{5.1.2 SDQN and SDQN-n Scheduler Experimental Data}}{5.1.2 SDQN and SDQN-n Scheduler Experimental Data}}\label{sdqn-and-sdqn-n-scheduler-experimental-data}}

In this scenario, 50 new pods were instantiated, each executing a no‑op workload to simulate compute‑intensive processing. Figure\,3 presents the results of the first test conducted with the SDQN Kubernetes scheduler.

Test 1, slave1=13, slave2=13, slave3=21, slave4=3. Cluster CPU average utilization 25.21\%

\includegraphics[width=5.33333in,height=2.24583in]{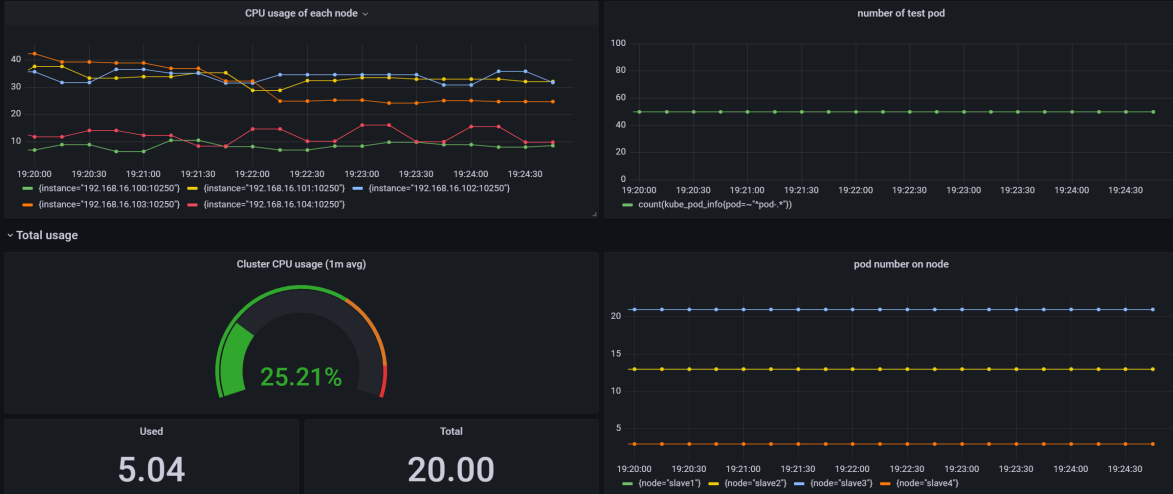}

Figure 3: SDQN scheduler grafana graph

The above figure is a Grafana example. To guard against the lack of representativeness in a single run, the following presents results from multiple experimental trials as Table 9.

Table 9: SDQN scheduler multiple data

\begin{table}[htbp]
\centering
\scriptsize
\setlength{\tabcolsep}{3pt}
\begin{tabular}{>{\raggedright\arraybackslash}p{0.20\linewidth} c c c c c c}
\toprule
\textbf{Scheduler Model} & \textbf{Trial} & \multicolumn{4}{c}{\textbf{Pod Distribution}} & \textbf{Average CPU Utilization} \\
\cmidrule(lr){3-6}
 &  & \textbf{Slave 1} & \textbf{Slave 2} & \textbf{Slave 3} & \textbf{Slave 4} &  \\
\midrule
\multirow{5}{*}{SDQN} & 1 & 13 & 13 & 21 & 3 & 25.21\% \\
 & 2 & 15 & 14 & 19 & 2 & 27.69\% \\
 & 3 & 16 & 14 & 19 & 1 & 26.39\% \\
 & 4 & 15 & 16 & 17 & 2 & 27.93\% \\
 & 5 & 20 & 11 & 16 & 3 & 28.84\% \\
\midrule
\textbf{Average} & \multicolumn{5}{c}{Coefficient of Variation (CV) =4.67\%} & 27.21\% \\
\bottomrule
\end{tabular}
\end{table}

In this scenario, 50 new pods were instantiated, each executing a no‑op workload to simulate compute‑intensive processing. Figure\,4 presents the results of the first test conducted with the SDQN-n(n=2) Kubernetes scheduler.Here, n=2 enforces that newly created pods are, as far as possible, concentrated on two nodes.

Test 1, slave1=25, slave2=25, slave3=0, slave4=0. Cluster CPU average utilization 21.01\%

\includegraphics[width=5.75417in,height=2.4625in]{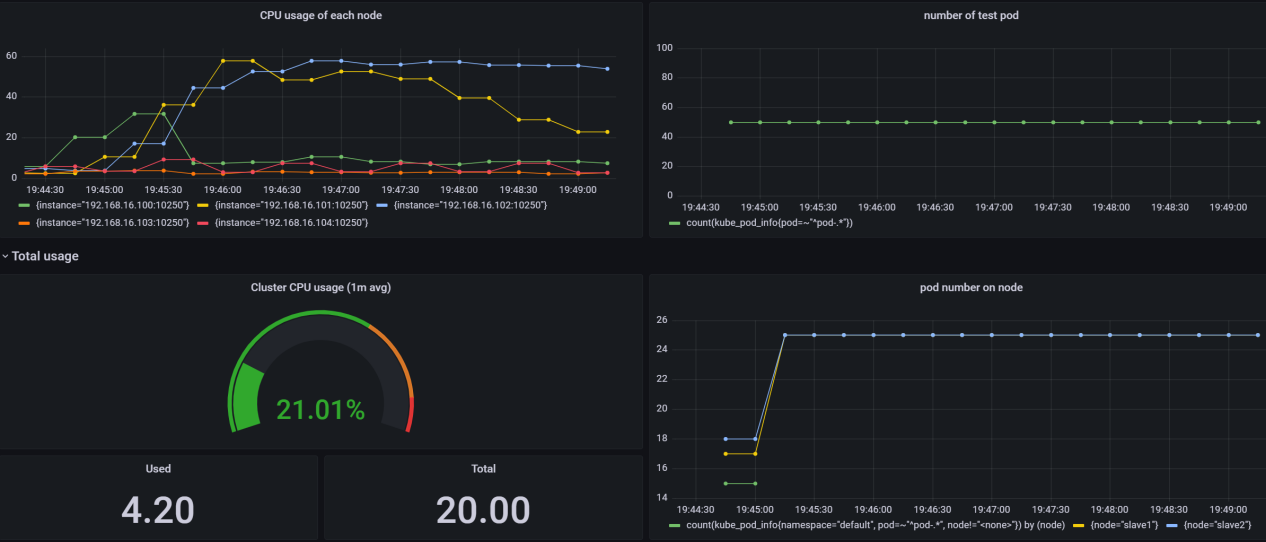}

Figure 4: SDQN-n scheduler grafana graph

The above figure is a Grafana example. To guard against the lack of representativeness in a single run, the following presents results from multiple experimental trials as Table 10.

Table 10: SDQN-n scheduler multiple data

\begin{table}[htbp]
\centering
\scriptsize
\setlength{\tabcolsep}{3pt}
\begin{tabular}{>{\raggedright\arraybackslash}p{0.20\linewidth} c c c c c c}
\toprule
\textbf{Scheduler Model} & \textbf{Trial} & \multicolumn{4}{c}{\textbf{Pod Distribution}} & \textbf{Average CPU Utilization} \\
\cmidrule(lr){3-6}
 &  & \textbf{Slave 1} & \textbf{Slave 2} & \textbf{Slave 3} & \textbf{Slave 4} &  \\
\midrule
\multirow{5}{*}{SDQN-n(n=2)} & 1 & 13 & 13 & 21 & 3 & 25.21\% \\
 & 2 & 2 & 2 & 23 & 23 & 22.57\% \\
 & 3 & 21 & 22 & 3 & 4 & 26.39\% \\
 & 4 & 25 & 25 & 0 & 0 & 22.01\% \\
 & 5 & 25 & 25 & 0 & 0 & 23.84\% \\
\midrule
\textbf{Average} & \multicolumn{5}{c}{Coefficient of Variation (CV) =4.00\%} & 22.35\% \\
\bottomrule
\end{tabular}
\end{table}

Although pod placement is enforced to concentrate new pods on two nodes whenever possible, if CPU utilization or other parameters exceed the reinforcement‑learning thresholds, a small number of pods will be scheduled on alternative nodes to prevent node failure and safeguard service continuity.

\hypertarget{lstmbased-and-transformerbased-scheduler-experimental-data}{%
\subsection{\texorpdfstring{\textbf{5.1.3 LSTM‑Based and} Transformer‑Based \textbf{Scheduler Experimental Data}}{5.1.3 LSTM‑Based and Transformer‑Based Scheduler Experimental Data}}\label{lstmbased-and-transformerbased-scheduler-experimental-data}}

In this scenario, 50 new pods were instantiated, each executing a no‑op workload to simulate compute‑intensive processing. Figure\,5 presents the results of the first test conducted with the LSTM‑Based Kubernetes scheduler.

\includegraphics[width=5.76181in,height=2.40833in]{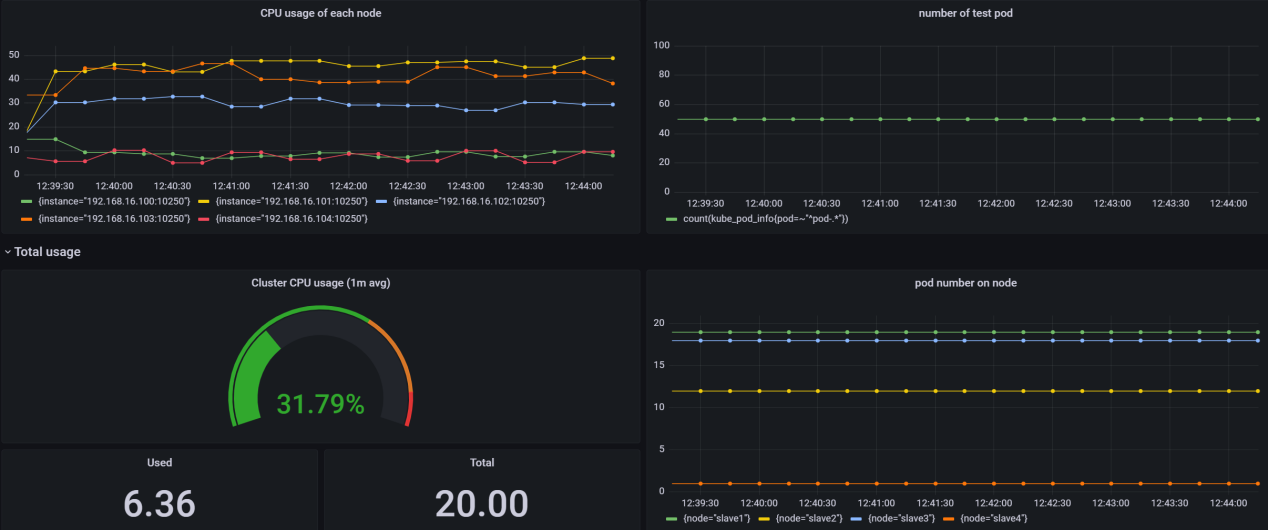}

Figure 5: LSTM‑Based scheduler grafana graph

The above figure is a Grafana example. To guard against the lack of representativeness in a single run, the following presents results from multiple experimental trials as Table 11.

Table 11: LSTM‑Based scheduler multiple data

\begin{table}[htbp]
\centering
\scriptsize
\setlength{\tabcolsep}{3pt}
\begin{tabular}{>{\raggedright\arraybackslash}p{0.20\linewidth} c c c c c c}
\toprule
\textbf{Scheduler Model} & \textbf{Trial} & \multicolumn{4}{c}{\textbf{Pod Distribution}} & \textbf{Average CPU Utilization} \\
\cmidrule(lr){3-6}
 &  & \textbf{Slave 1} & \textbf{Slave 2} & \textbf{Slave 3} & \textbf{Slave 4} &  \\
\midrule
\multirow{5}{*}{LSTM-Based} & 1 & 19 & 12 & 18 & 1 & 31.97\% \\
 & 2 & 19 & 11 & 19 & 1 & 32.87\% \\
 & 3 & 20 & 11 & 18 & 1 & 28.43\% \\
 & 4 & 15 & 16 & 17 & 2 & 29.73\% \\
 & 5 & 16 & 15 & 17 & 2 & 29.67\% \\
\midrule
\textbf{Average} & \multicolumn{5}{c}{Coefficient of Variation (CV) =5.35\%} & 30.53\% \\
\bottomrule
\end{tabular}
\end{table}

The results of the Transformer‑based scheduler closely paralleled those of the LSTM‑based scheduler; data from multiple runs are presented in Table \,12.

Table 12: Transformer‑Based scheduler multiple data

\begin{table}[htbp]
\centering
\scriptsize
\setlength{\tabcolsep}{3pt}
\begin{tabular}{>{\raggedright\arraybackslash}p{0.20\linewidth} c c c c c c}
\toprule
\textbf{Scheduler Model} & \textbf{Trial} & \multicolumn{4}{c}{\textbf{Pod Distribution}} & \textbf{Average CPU Utilization} \\
\cmidrule(lr){3-6}
 &  & \textbf{Slave 1} & \textbf{Slave 2} & \textbf{Slave 3} & \textbf{Slave 4} &  \\
\midrule
\multirow{5}{*}{Transformer-Based} & 1 & 20 & 11 & 18 & 1 & 29.25\% \\
 & 2 & 19 & 11 & 19 & 1 & 30.01\% \\
 & 3 & 19 & 12 & 18 & 1 & 30.48\% \\
 & 4 & 15 & 16 & 17 & 2 & 30.47\% \\
 & 5 & 15 & 16 & 17 & 2 & 30.52\% \\
\midrule
\textbf{Average} & \multicolumn{5}{c}{Coefficient of Variation (CV) =1.61\%} & 30.15\% \\
\bottomrule
\end{tabular}
\end{table}

\includegraphics[width=6.49375in,height=3.45208in]{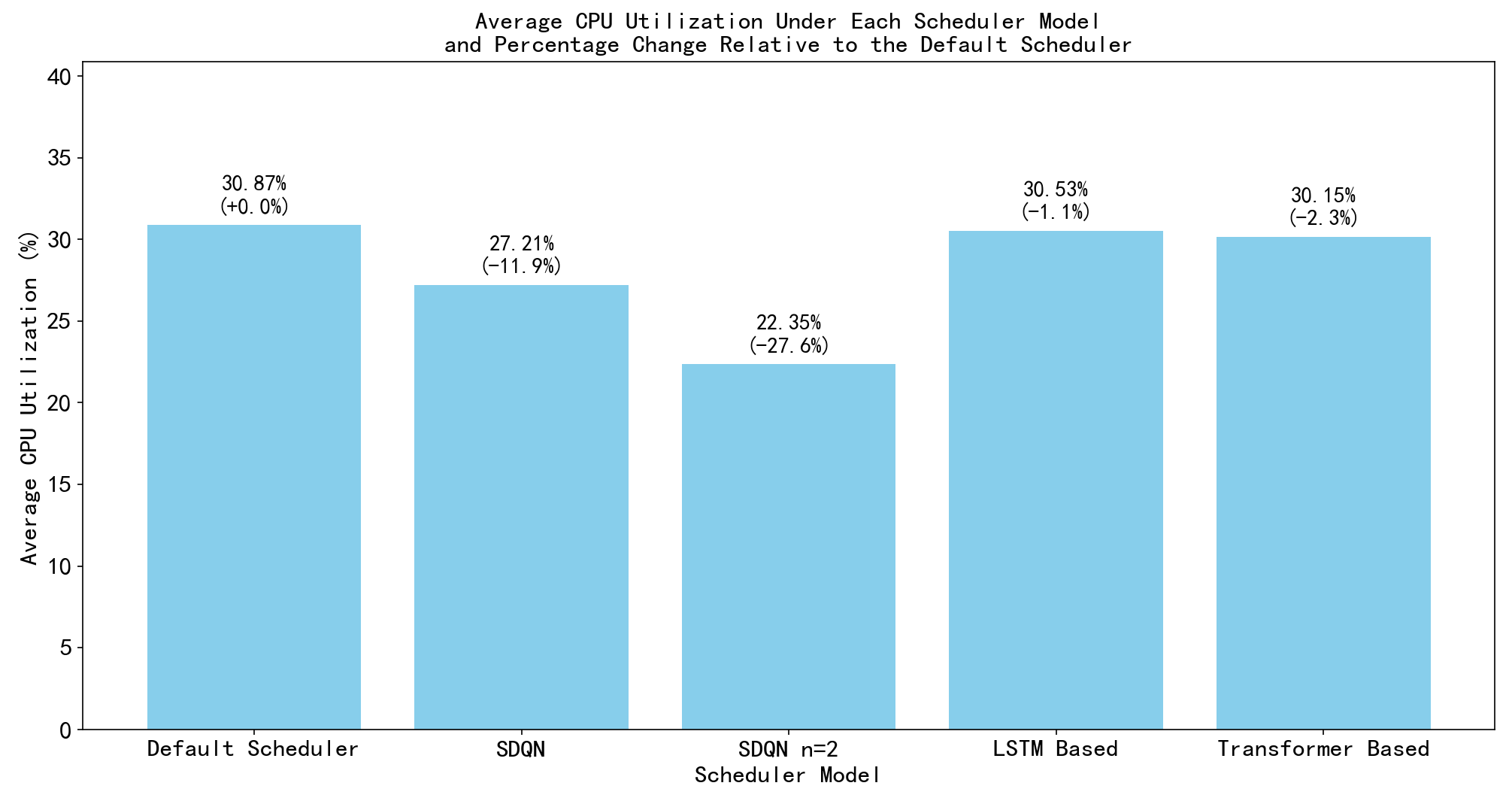}

\textbf{Figure 6:} Comparison of different algorithms

The experimental results as Figure 6 demonstrate that the default scheduler underperforms SDQN in compute‑intensive scenarios, and that SDQN‑n likewise achieves superior performance compared with the default scheduler. Although Kubernetes typically aims to distribute pods evenly across all available nodes, concentrating compute‑intensive pods on two nodes yields the lowest average CPU utilization in this study, suggesting that, when resources are abundant, scaling down under‑utilized nodes can advance green data‑center objectives. In contrast, the custom LSTM‑based and Transformer‑based schedulers exhibit average CPU utilizations comparable to the default scheduler, showing no significant advantage. The superior effectiveness of SDQN arises from reinforcement learning's ability to adapt to each node real‑time state, placing pods in optimally suited locations and thereby reducing overall CPU usage.

\hypertarget{conclusions-and-future-work}{%
\subsection{\texorpdfstring{\textbf{6} Conclusions and Future Work}{6 Conclusions and Future Work}}\label{conclusions-and-future-work}}

\textbf{6.1 Conclusions}

\begin{itemize}
\item
  1. Adapt the current models to outperform the default scheduler across a broader range of workload types and cluster configurations.
\item
  2. Investigate the SDQN‑n consolidation strategy as a blueprint for energy‑efficient (``green'') data centers and clusters.
\item
  3. Refine and tune the hyperparameters of SDQN and SDQN‑n to further enhance resource savings and scheduling robustness.

  \textbf{6.2 Future Work}
\item
  1. Adapt the current models to outperform the default scheduler across a broader range of workload types and cluster configurations.
\item
  2. Investigate the SDQN‑n consolidation strategy as a blueprint for energy‑efficient (``green'') data centers and clusters.
\item
  3. Refine and tune the hyperparameters of SDQN and SDQN‑n to further enhance resource savings and scheduling robustness.
\end{itemize}

1. SIMON, M., HURAJ, L., and BúCIK, N., 2023. A Comparative Analysis of High Availability for Linux Container Infrastructures. \emph{Future Internet 15}, 8 (Aug). DOI= \url{http://dx.doi.org/ARTN} 253

10.3390/fi15080253.

2. SILVA, S.N., GOLDBARG, M.A.S.D., DA SILVA, L.M.D., and FERNANDES, M.A.C., 2024. Application of Fuzzy Logic for Horizontal Scaling in Kubernetes Environments within the Context of Edge Computing. \emph{Future Internet 16}, 9 (Sep). DOI= \url{http://dx.doi.org/ARTN} 316

10.3390/fi16090316.

3. BARESI, L., HU, D.Y.X., QUATTROCCHI, G., and TERRACCIANO, L., 2021. : Vertical and Horizontal Resource Autoscaling for Kubernetes. \emph{Service-Oriented Computing (Icsoc 2021) 13121}, 821-829. DOI= \url{http://dx.doi.org/10.1007/978-3-030-91431-8_59}.

4. VIGLIANISI, E., DALLAGO, M., and CECCATO, M., 2020. RESTTESTGEN: Automated Black-Box Testing of RESTful APIs. \emph{2020 Ieee 13th International Conference on Software Testing, Validation and Verification (Icst 2020)}, 142-152. DOI= \url{http://dx.doi.org/10.1109/Icst46399.2020.00024}.

5. XIAO, Z.J. and HU, S., 2022. DScaler: A Horizontal Autoscaler of Microservice Based on Deep Reinforcement Learning. \emph{2022 23rd Asia-Pacific Network Operations and Management Symposium (Apnoms 2022)}, 121-126.

6. LAI, W.-K., WANG, Y.-C., and WEI, S.-C.J.I.I.O.T.J., 2023. Delay-aware container scheduling in kubernetes \emph{10}, 13, 11813-11824.

7. DEHNASHI, M.N., MOMTAZPOUR, M., and JAVADI, S.A., 2024. A Communication-Aware Scheduler for Containers in a Kubernetes Environment Using Girvan-Newman Clustering. In \emph{2024 32nd International Conference on Electrical Engineering (ICEE)} IEEE, 1-5.

8. KIM, E., LEE, K., and YOO, C.J.T.J.O.S., 2023. Network SLO-aware container scheduling in Kubernetes \emph{79}, 10, 11478-11494.

9. WANG, Z., ZHANG, X., and YANG, L., 2023. EIS: Edge Information-Aware Scheduler for Containerized IoT Applications. In \emph{2023 IEEE International Conference on Edge Computing and Communications (EDGE)} IEEE, 280-289.

10. SAITO, D., HU, S., and SATO, Y., 2024. A Microservice Scheduler for Heterogeneous Resources on Edge-Cloud Computing Continuum. \emph{2024 Ieee Symposium in Low-Power and High-Speed Chips, Cool Chips 27}. DOI= \url{http://dx.doi.org/10.1109/Coolchips61292.2024.10531183}.

11. HUI, A.N.J. and LEE, B.S., 2021. Epsilon: A microservices based distributed scheduler for kubernetes cluster. In \emph{2021 18th International Joint Conference on Computer Science and Software Engineering (JCSSE)} IEEE, 1-6.

12. QIAO, Y., XIONG, J., and ZHAO, Y.J.C.C., 2025. Network-aware container scheduling in edge computing \emph{28}, 2, 78.

13. JIAN, Z.L., XIE, X.S., FANG, Y.Z., JIANG, Y.B., LU, Y., DASH, A., LI, T., and WANG, G.L., 2024. DRS: A deep reinforcement learning enhanced Kubernetes scheduler for microservice-based system. \emph{Software-Practice \& Experience 54}, 10 (Oct), 2102-2126. DOI= \url{http://dx.doi.org/10.1002/spe.3284}.

14. JORGE-MARTINEZ, D., BUTT, S.A., ONYEMA, E.M., CHAKRABORTY, C., SHAHEEN, Q., DE-LA-HOZ-FRANCO, E., ARIZA-COLPAS, P.J.I.J.O.S.A.E., and MANAGEMENT, 2021. Artificial intelligence-based Kubernetes container for scheduling nodes of energy composition, 1-9.

15. WANG, X., ZHAO, K., and QIN, B.J.M., 2023. Optimization of task-scheduling strategy in edge kubernetes clusters based on deep reinforcement learning \emph{11}, 20, 4269.

16. CHENG, W., XU, Y., XU, Q., ZHANG, H., LI, X., and SHAO, X., 2023. CSFRL: A Reinforcement Learning Technology Enabled Computing Power Scheduling Framework Based on Kubernetes. In \emph{2023 IEEE 34th Annual International Symposium on Personal, Indoor and Mobile Radio Communications (PIMRC)} IEEE, 1-6.

17. ROTHMAN, J. and CHAMANARA, J., 2023. An RL-Based Model for Optimized Kubernetes Scheduling. In \emph{2023 IEEE 31st International Conference on Network Protocols (ICNP)} IEEE, 1-6.

18. REJIBA, Z. and CHAMANARA, J.J.A.C.S., 2022. Custom scheduling in kubernetes: A survey on common problems and solution approaches \emph{55}, 7, 1-37.

\end{document}